\documentstyle[12pt,aasms4]{article}
\renewcommand{\arraystretch}{0.3}

\begin{document}

\title{Propagation of UHECRs from the Sources in \\
the Super-Galactic Plane}
\author{Yoshinori Ide\altaffilmark{1}, Shigehiro Nagataki\altaffilmark{1}, Sinya Tsubaki\altaffilmark{1},\\
 Hiroyuki Yoshiguchi\altaffilmark{1} and Katsuhiko Sato\altaffilmark{1,2}}
\noindent
\altaffilmark{1}{Department of Physics, School of Science, the University
of Tokyo, 7-3-1 Hongo, Bunkyoku, Tokyo 113-0033, Japan}\\
\altaffilmark{2}{Research Center for the Early Universe, School of
Science, the University of Tokyo, 7-3-1 Hongo, Bunkyoku, Tokyo 113-0033, Japan} \\

\begin{abstract}
We have performed the detailed numerical simulations on the propagation of
the UHE protons in the energy range $E=(10^{19.5} - 10^{22.0}$) eV in
the relatively strong extra-galactic magnetic field with strength
$B= (10, 100)$ nG within about 40 Mpc. In this case, the deflection
angles of UHECRs become so large that the no counterparts problem is simply
solved. As for the source distribution, we assumed that it is proportional
to the number distribution of galaxies within the GZK sphere.
We have found many clusters, which mean the small-scale anisotropy,
in our simulations. It has been also shown that the observed
energy spectrum is well reproduced in our models without any fine-tuned
parameter. We have used the correlation value in order to investigate
statistically the similarity between the distribution of arrival directions
of UHECRs and that of galaxies. We have found that each correlation value
for each parameter set begins to converge when the number of the detected
events becomes ${\large O}(10^3)$. Since the expected number counts by the
experiment of the next generation such as TA, HiRes, Auger, and EUSO are
thought to be the order of $10^3$, we will be able to determine the source
distribution and values of the parameters in this study in the very near
future. Compared with the AGASA data, the significant anisotropy on the
arrival directions of UHECRs are found in the analysis of first and second
harmonics. This may originate from the incompleteness of the ORS database.
This problem may be also solved if the source distribution is slightly
changed. For example, this problem may be solved if we assume that
UHECRs come from some of the galaxies such as AGNs and radio galaxies.
\end{abstract}

\section{INTRODUCTION} \label{intro}
\indent

The observed differential cosmic ray spectrum is remarkably featureless
and extends beyond $10^{20}$ eV (\cite{takeda99}). So far, the observed
number of the Ultra High Energy Cosmic Rays (hereafter UHECRs) whose
energies are
above $10^{20}$ eV (this is the definition of UHECRs in this study)
is only 25 (\cite{virmani00}). On the other
hand, a new generation of the ground-based large aperture experiments
such as Telescope Array (hereafter TA; \cite{ta00}), HiRes
(\cite{wilkinson99}) and South and North Auger (\cite{capelle98}) is expected
to detect about 1000 UHECRs
until 2010 (\cite{zas01}). Moreover, the mission proposed as EUSO is devoted
to the exploration of UHECRs from satellites and is expected to detect
about 1000 UHECRs per year (\cite{bhattacharjee98}). We will be able to discuss
statistically the feature of UHECRs, such as arrival direction, in the
very near future.

The mechanism to produce such highly energetic cosmic rays is still
controversial.
In most of the conventional acceleration scenarios, which are called bottom-up
scenarios, effects of diffusive shock acceleration are taken into
consideration (e.g., \cite{biermann95}; \cite{halzen97}; \cite{waxman00}).
So far, by using
Hillas-plot (\cite{hillas84}; \cite{selvon00}),
gamma-ray bursts (GRBs) and/or active galactic nuclei (AGN) are considered
as probable candidates. 
On the other hand, there is a large number of production models based on
exotic particle physics scenarios (see \cite{bhattacharjee00} and references
therein). Sometimes effects of interaction or
collapse of Topological Defects are taken into consideration to
produce such massive exotic particles (e.g., \cite{bhattacharjee92}).
These scenarios
are called top-down scenarios. In this study, we mainly consider bottom-up
scenarios although top-down scenarios are very attractive and important.
This is because top-down scenarios are highly model-dependent and
observational constraints on them are little at present. Thus we investigate
in this study whether the present observations on UHECRs can be explained by
bottom-up scenarios or not.

In the bottom-up scenarios, charged particles are accelerated due to
Fermi acceleration mechanism and obey a power law spectrum (\cite{fermi49}).
However,
the large distances between the Earth and potential UHECRs sources like
GRBs and AGNs lead to another problem called GZK effect (\cite{greisen66};
\cite{zatsepin66}).
As for the protons, the energy at which the GZK cutoff takes place
($\sim 7 \times 10^{19}$ eV) is given by the threshold for photopion
production in the collisions of protons and CMB photons. 
It is reported that the loss length ($l_{\rm loss} = Edl/dE$) falls below
13 Mpc above 100 EeV (\cite{stanev00}). As for nuclei, the situation
is considered to be worse due to the photo-disintegration
mechanism (\cite{stanev00}). Thus
we assume in this study that the composition of UHECRs is proton.

Taking the GZK sphere for protons into consideration, we can easily
understand the difficulty of the situation, that is, no plausible
source counterparts within the GZK sphere have been found within
a few degrees from the arrival directions of UHECRs, which are considered
to be the typical deflection angles of UHECRs (e.g., \cite{blasi98}). 
Moreover, there is a puzzling problem on the distribution of the
arrival directions of UHECRs (e.g., \cite{blasi98}). It is reported that there
is no significant large-scale anisotropy of arrival direction distribution
of UHECRs. On the other hand, there
are one triplet and three doublets within a separation angle of 2.5 degrees
for the 47 cosmic rays above $4 \times 10^{19}$ eV, and the probability of
observing these clusters by a chance coincidence under an isotropic
distribution is smaller than 1$\%$ (\cite{takeda99}).
We have to give a natural explanation for these puzzling problems.

In this paper, we propose a bottom-up scenario in which the sources
of UHECRs are correlated with the Super-Galactic Plane (SGP).
This is because the probability of observing the clusters by a
chance coincidence under such a distribution becomes relatively
large ($\sim 10$ $\%$) (\cite{lemoine99}). In fact, we show in this
study that such clusters emerge very frequently in our model. It is also shown
that the observed energy spectrum is well reproduced in our models without
any fine-tuned parameter. Moreover, we show statistically
that the correlation between the arrival directions of UHECRs and
SGP can not be determined significantly at the present number of
data (25 enevts), which is consistent with the AGASA data (\cite{takeda99}).
We conclude that about 1000 events, which are attained in the very
near future as mentioned above, are necessary to determine whether
the source distribution is correlated with SGP or not.
Moreover, we introduce a relatively large amplitude of the extra-galactic
magnetic field ($\sim 10$ nG) in order to solve the no counterpart
problem. This is our picture for solving the problems on UHECRs.

In this study, images of arrival directions of UHECRs are simulated.
Optical Redshift Survey (ORS) data (\cite{santiago95}) is used as the source
distribution of UHECRs. Such a realistic data has not been used in the
simulation of propagation of UHECRs in the previous works.
Inhomogeneity of the arrival directions of UHECRs are also discussed
statistically by introducing the correlation value, which has not
been introduced in the previous works (e.g. \cite{lemoine99}).
These are new points of this work. We also show one realization of
arrival directions of UHECRs above 4$\times 10^{19}$ eV in order to
compare our results with the AGASA data (\cite{takeda99}). 
First and second harmonics analysis for our model is also presented in
order to compare directly our results with the AGASA data (\cite{takeda99}).
Our conclusion has been already stated above.

In section~\ref{model}, we show our method of calculation. 
Results are shown in section~\ref{result}. Summary and discussion
are presented in section~\ref{summary}.

\section{METHOD OF CALCULATION} \label{model}
\subsection{Method of Calculation for Propagation of UHECRs}
\label{propagation}
\indent

In this subsection, we describe the method of Monte Carlo simulations
for the
propagation of UHECRs. At first, we assume that the composition of
UHECRs is proton. We also assume that the initial energy spectrum
of the UHECRs obeys the power-law, that is, $dN/dE \propto E^{-2}$,
where $N$ denotes the number of UHECRs. Initial energies of UHECRs
are assumed to be in the range of ($10^{19.5}$ -- $10^{22}$)eV.
The number of injected UHECRs in a simulation is $10^6$ except for
the cases of $l_c$ = 1Mpc, where $l_c$ is the correlation length of
the extra-magnetic field and is explained below. In the case of
($B$, $l_c$) = (10nG, 1Mpc), the number of the injected particles is
$10^5$, where $B$ is the strength of the extra-magnetic field and
is explained below. In the case of ($B$, $l_c$) = (100nG, 1Mpc),
the number of injected particles is $2.5 \times 10^4$.

As for the energy loss processes, electron-positron pair creation
and photopion production in the CMB field are included.
Particles below $\sim 10^{19.5}$
eV lose their energies mainly by pair creations and above it lose their
energies mainly by photopion production (\cite{yoshida93}).
We adopt the formulation of the energy loss rate for the pair production on
isotropic photons, which has been shown by Chododowski et al. (1992)
According to them, the energy loss rate of a relativistic nucleus for
the pair production on isotropic photons is given by
\begin{equation}  
- \frac{d\gamma}{dt} = \alpha r^2_0 c Z^2 \frac{m_e}{m_A}\int^{\infty}_2
d\kappa n \left( \frac{\kappa}{2\gamma}\right)
\frac{\varphi(\kappa)}{\kappa ^2}, 
\label{eqn1}
\end{equation}
where $\gamma$ is the Lorenz factor of the particle, $\kappa =  2k
\gamma$ ( $k$ is the momentum of the particle in units of $m_ec$),
$n(\kappa)$, $\alpha$, $r_0$, $Z$, and $m_A$ are the photon distribution
in the momentum space, the fine-structure constant, the classical electron
radius, the charge of the particle, and the rest mass of the particle,
respectively. 
For the energy range $E \geq 10^{19.0}$ eV, $\varphi$ can be represented
well as
\begin{equation}
\varphi \rightarrow \kappa \sum ^{3} _{i=0} d_i \ln ^i \kappa,
\label{eqn2}
\end{equation}
\begin{equation} 
d_0 \simeq -86.07,\; d_1 \simeq 50.96,\; d_2 \simeq -14.45,\;
d_3=8/3.
\label{eqn3}
\end{equation}
We also adopt the formulation of the energy loss rate for the photopion
production on isotropic photons, which has been shown by Achterberg et
al. (1999). According to them, the interaction length of the photopion
production, $l_{p\gamma}(E_p)$, in the CMB can be written as
\begin{equation}
l_{p\gamma}(E_p)/ \mbox{Mpc} \simeq \left \{
\begin{array}{ll}
0.9\left( \frac{E_p}{E_{\mbox{{\scriptsize b}}}}\right) ^2
e^{E_{\mbox{{\scriptsize b}}}/E_p}, &\quad \mbox{for}\quad E_p \leq
0.2 E_{\mbox{{\scriptsize b}}} \\
4.8, & \quad \mbox{for}\quad  E_p > 0.2 E_{\mbox{{\scriptsize b}}} 
\end{array}
\right.,
\label{eqn4}
\end{equation}
where $E_p$ is the proton energy in the cosmic rest frame and
$E_{\mbox{{\scriptsize b}}}$ is defined as
\begin{equation}
E_{\mbox{{\scriptsize b}}} \equiv \frac{m_p \varepsilon _{\mbox{{\scriptsize
th}}}}{2  k_{\mbox{{\scriptsize b}}}T } \left( = E_p
\frac{\varepsilon _{\mbox{{\scriptsize th}}}}{2 \gamma _p  
k_{\mbox{{\scriptsize b}}}T} \right).
\label{eqn5}
\end{equation}
Here $\varepsilon _{\mbox{{\scriptsize th}}} \equiv m_{\pi} (1+m_{\pi}/2m_p)
\simeq 145$MeV is the threshold energy of photon in the proton rest frame.
As for the mean inelasticity, it can be written as
\begin{equation}
K_p \equiv \frac{E_t ^2 + m_{\pi} ^2-m_p^2}{2E^2_t},
\label{eqn6}
\end{equation}
where $E_t$ is defined as
\begin{equation}
E_t = \sqrt{ m_p^2+2m_p\varepsilon _0}. 
\label{eqn7}
\end{equation}
Here $\varepsilon _0$ is the photon energy in the proton rest frame.
As for the maximum spread of inelasticity from the mean one,
it can be written as
\begin{eqnarray}
K' &\equiv& \frac{ \sqrt{(E_t^2-m_+^2)(E_t^2-m_-^2)} }{2E^2_t}  \\
&=& \sqrt{(K_P+K_{+})(K_P-K_{-})}, 
\label{eqn8}
\end{eqnarray}
where $m_{\pm} \equiv m_p \pm m_{\pi}$ and $K_{\pm} \equiv m_{\pi} 
/(m_p \mp m_{\pi})$. 
Since photopion productions obey the Poisson statistics, we can calculate
the energy losses of UHECRs by the photopion production using the
formulations mentioned above.

Next, the details on the extra-magnetic field, which is little known
theoretically and observationally, is
described. Its strength and detailed structure are not known, although
the strength of the magnetic field in the local super cluster is thought
to be relatively strong.
Only its observational upper limit, $\sim 1 \mu$G in the meaning
of r.m.s., has been obtained from the Faraday rotation of the
distant sources (\cite{ryu98}).
In this study, 10 nG and 100nG are adopted as the r.m.s. of the magnetic
field. They seems to be strong to be sure (\cite{sigl98}). However, such
strong magnetic field is required to solve the problem that there seems
to be no plausible source counterparts within the GZK sphere within
a few degrees from the arrival directions of UHECRs.
We also assume that the magnetic field is represented as the Gaussian
random field with zero mean and a power-law spectrum.
Thus, $\langle B^2(k)\rangle$ can be written as
\begin{eqnarray}
\langle B^2(k)\rangle && \propto k^{n_H} \;\;\; {\rm for} \;\;\; k \le
k_{\mbox{{\scriptsize cut}}} \\
&& =0 \;\;\;\;\;\;\; \rm otherwise,
\label{eqn9}
\end{eqnarray}
where $k_{\mbox{{\scriptsize cut}}} = 2\pi /l_{\mbox{{\scriptsize
cut}}} = 16 \pi /l_{\rm c} $ characterizes the numerical cut-off scale
that is explained below and $n_H$ is chosen to be -11/3 so as to represent
the Kolmogorov spectrum. As for the wave numbers of the magnetic field,
8 discrete ones are introduced in this study. In practice, $k_i=
0, \pm k_0, \pm 2k_0, \pm 3k_0, 4k_0$ are used as wave numbers. Here
$k_0$ is $2\pi /l_c$. As for the correlation length of the magnetic
field, $l_c$,
three different values are adopted in this study. They are 1 Mpc, 10 Mpc,
and 40 Mpc. 1 Mpc is widely used as the typical value of the correlation
length of the extra-galactic magnetic field. 10 Mpc and 40 Mpc represent
the scale hight and the scale length of the local super cluster, respectively. 
We separate one cubic cell of the size $l_c$ into 512 (= $8^3$) smaller
cells of the size $l_c/8$, which corresponds to $l_{\rm cut}$ mentioned above.
The magnetic field are assigned to each small cell. 
As for the boundary condition, the periodic
boundary condition is adopted in order to reduce storage data for magnetic
field components.

Finally, we explain the source distribution of UHECRs.
In this study, we assume that the distribution of sources of UHECRs
is proportional to that of the galaxies within the GZK sphere.
In practice, we use the realistic data from the Optical Redshift Survey
(ORS; \cite{santiago95}). It is noted that the source
distribution will be such like that when GRBs occur very frequently in
every galaxies. It is also noted that the source distribution may be
such like that in the top-down scenarios, too. This is because the sources
of UHECRs such as massive exotic particles will be trapped in the
gravitational potential of galaxies and/or clusters of galaxies.
As for the dependence of our conclusion presented in this study on the
source distribution, we will report it in the forthcoming paper.

\subsection{Statistics on the Arrival Directions of UHECRs}\label{statics}
\subsubsection{Method of Calculation of Correlation Value}\label{correlation}
\indent

In this study, we introduce the correlation value in order to investigate
statistically the similarity between the distribution of arrival directions
of UHECRs and that of galaxies within the GZK sphere.
The correlation value, $\mit\Xi$, between two distributions $f_i$ and
$f_s$, is defined as
\begin{equation}
\mit\Xi (f_i,f_s) \equiv \frac{\rho (f_i,f_s)}{\sqrt{\rho (f_i,f_i)
\rho (f_s,f_s)}},
\label{eqn10}
\end{equation}
where
\begin{equation}
\rho (f_a,f_b) \equiv \sum _{j,k}\left( \frac{f_a(j,k)-\bar f_a}{\bar f_a }
\right)\left( \frac{f_b(j,k)-\bar f_b }{\bar f_b } \right)  \frac{\Delta
\Omega (j,k)}{4\pi}.
\label{eqn11} 
\end{equation}
Here subscripts $j$ and $k$ discriminate each cell of the sky,
$\Delta\Omega (j,k)$ denotes the solid angle of the ($j,k$) cell, and
$\bar f$ means the average of $f$. In this study, the size of the cell
is chosen to be $1^{\circ} \times 1^{\circ}$.
The meaning of $\mit\Xi$ is as follows. By definition, $\mit\Xi$ ranges
from $-1$ to $+1$. When $\mit\Xi =$ +1, two distributions
are same exactly. When $\mit\Xi = $ -1, two distributions are exactly
opposite. When $\mit\Xi = $ 0, we can not find any resemblance between two
distributions.

Strictly speaking, the angular images obtained by numerical simulations
do not mean the exact distributions of the UHECRs
which we will detect indeed, but the probability density distributions for
the arrival directions of UHECRs. Thus, in order to investigate the
correlation between the results of numerical simulations and the distribution
of galaxies in the ORS catalogue statistically, we have to make a lot of
data samples for the arrival
directions of UHECRs assuming the numerically obtained probability
density distribution.
In order to estimate the effects of selecting the events stochastically,
the number of data samples is set to be 1000 for the same number of events
of UHECRs, $N$, under the same condition (e.g., $B$ and $l_c$). Hence,
a mean value and a standard deviation of $\mit\Xi$s can be obtained for
each $N$ and condition.

In practice, we choose the 6 different values for the number ($N$) of events of
UHECRs. These are 25, $10^2$, 320, $10^3$, 3200, and $10^4$. 25 is the present
number of events of UHECRs (\cite{virmani00}). Other values are chosen
arbitrarily only paying attention to the fact that the expected number
counts detected by the experiment of the next generation such as TA, HiRes,
Auger, and EUSO are thought to be the order of $10^3$
(\cite{bhattacharjee98}).

\subsubsection{Analysis of First and Second Harmonics}\label{harmonics}
\indent

In order to search for the global anisotropy in the arrival directions of
UHECRs, we apply harmonics analysis to the galactic longitude distribution of
events (\cite{hayashida99}). It should be noted that we do not investigate
the right ascension distribution but investigate the
galactic longitude distribution so as not to suffer from the problem of
the incompleteness of the ORS catalog, which
does not contain any data within $| b | \le 20^{\circ}$ (see
Figure~\ref{fig5}). We will explain the definition of it below
(see Hayashida et al. 1999 in detail). The $m$-th harmonic amplitude,
$r$, and phase of maximum, $\theta$, are obtained for a sample of $n$
measurements of phase, $\phi_1$, $\phi_2$, $\cdot \cdot \cdot$, $\phi_n$
(0 $\le \phi_i \le 2 \pi$) from:
\begin{equation}
r = (a^2 + b^2)^{1/2}
\label{eqn121}
\end{equation}
\begin{equation}
\theta = \tan ^{-1} (b/a)
\label{eqn12}
\end{equation}
where, $a = \frac{2}{n} \Sigma_{i = 1}^{n} \cos m \phi_i  $,
$b = \frac{2}{n} \Sigma_{i = 1}^{n} \sin m \phi_i  $.

The following $k$ represents the statistical significance. If events with
total number $n$ are uniformly distributed in galactic longitude, the chance
probability of observing the amplitude $\ge r$ is given by,
\begin{equation}
P = \exp (-k),
\label{eqn13}
\end{equation}
where
\begin{equation}
k = n r^2/4.
\label{eqn14}
\end{equation}
We take $n$ to be 47 in the range $E \ge 4 \times 10^{19}$eV in order
to compare our results with AGASA data (\cite{takeda99}), although the right
ascension distribution of events are investigated in their harmonics analysis.
We also take $n$ to be 100 in each energy bin in the range $10^{19.15}
\le E \le 10^{19.5}$eV. However, it is noted that the results of harmonic
analysis for the energy range $E \le 10^{19.5}$eV may not be reliable.
This is because the initial energy of UHECRs is restricted to be in the
range of $10^{19.5} \le E \le 10^{22}$eV to save the CPU time.

\section{RESULTS} \label{result}
\subsection{Arrival Directions and Energy Spectra of UHECRs}\label{Images}
\indent

In this subsection, we show the images of the distribution of the
arrival directions of
UHECRs. We note again that the resolution of the image, that is,
the size of the cell is taken to be $1^{\circ} \times 1^{\circ} $.
It is also noted that these images show the probability density
distributions of arrival directions of UHECRs. The observed distribution
should be interpreted as one realization. That is why we have to discuss the
correlation among them statistically. In the following subsections, the
statistical discussion is presented.

\begin{figure}
\epsscale{1.0}
\plottwo{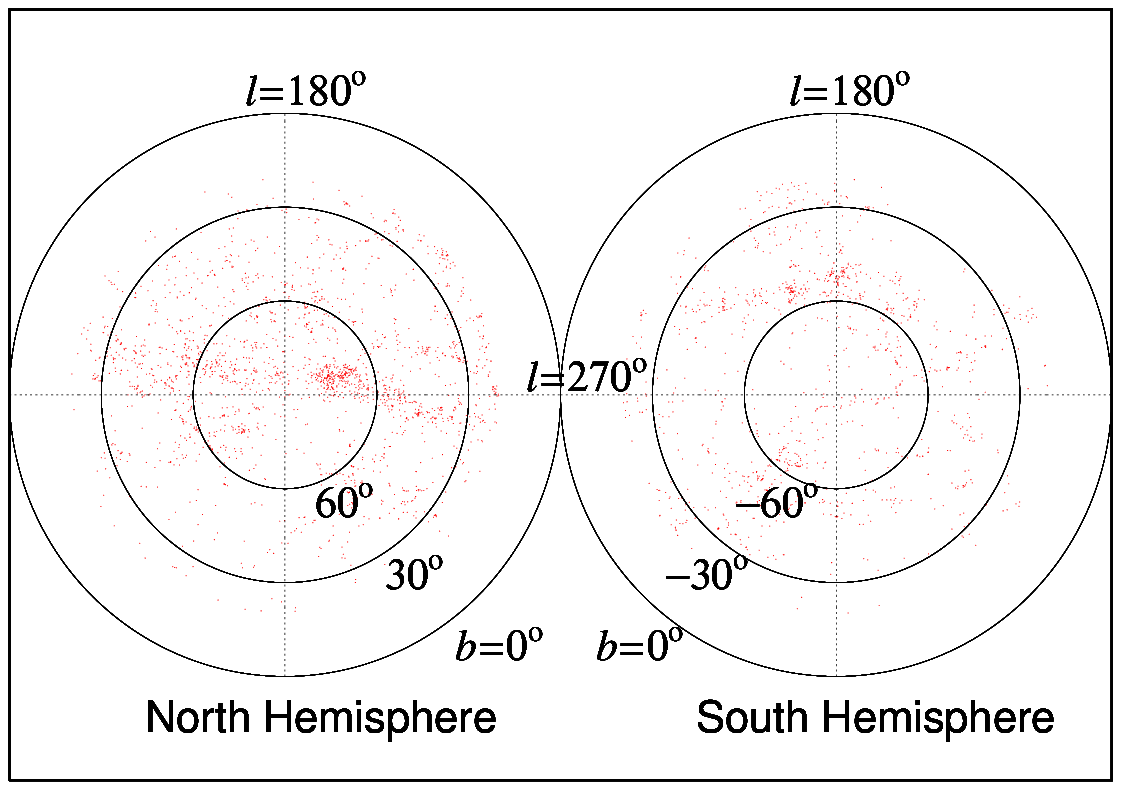}{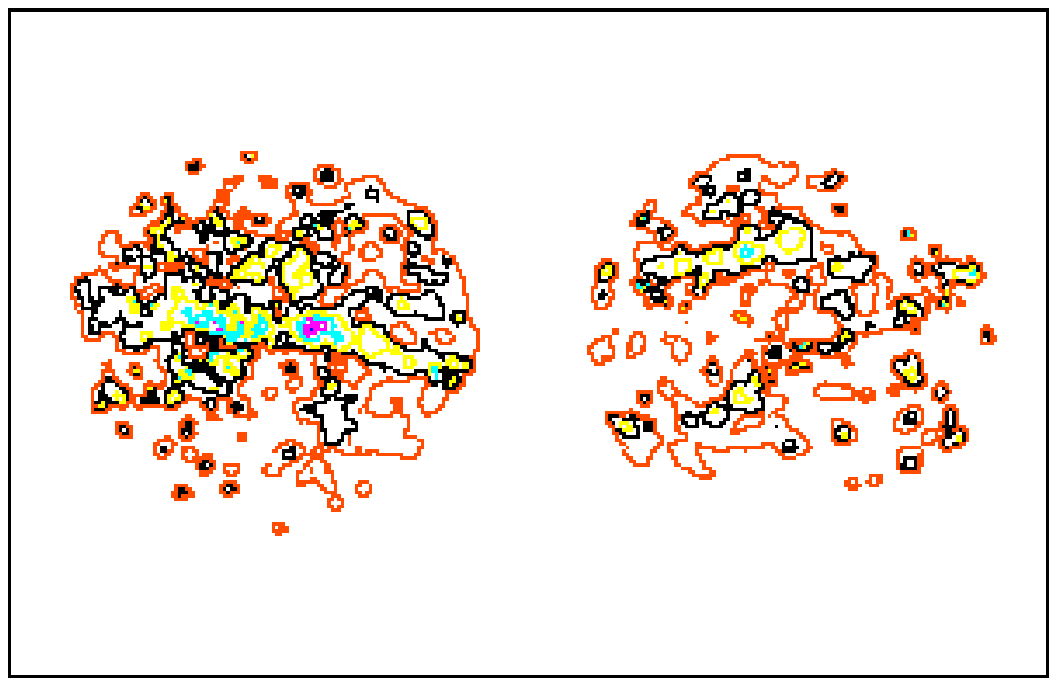}
\figcaption{
(a) left panel: angular distribution of galaxies in Optical Redshift
Survey (ORS) data within 40Mpc. (b) right panel: distribution of arrival
directions of
UHECRs in the case of $B=10$ nG, $l_c=10$ Mpc, $H_0=75$ km/s/Mpc, and
$E \geq 10^{19.5}$ eV. Galactic coordinate is used and left parts correspond
to the north galactic hemisphere. The resolution of the image is set to
be $1^{\circ} \times 1^{\circ} $. The inter-contour interval is 0.5 in the
logarithm to base 10 of the integral flux per solid angle.
\label{fig1}}
\end{figure}

At first, in the left panel of Figure~\ref{fig1},
the distribution of the
galaxies within 40 Mpc in ORS data catalogue is shown.
In the right panel of Figure~\ref{fig1},
the image in the case of $B=10$ nG, $l_c=10$ Mpc, $H_0=75$ km/s/Mpc, and
$E \geq 10^{19.5}$ eV is shown. The inter-contour interval is 0.5 in the
logarithm to base 10 of the integral flux per solid angle.
We can find that the image is distorted
due to the deflection caused by extra-galactic magnetic field.

\begin{figure}
\epsscale{1.0}
\plottwo{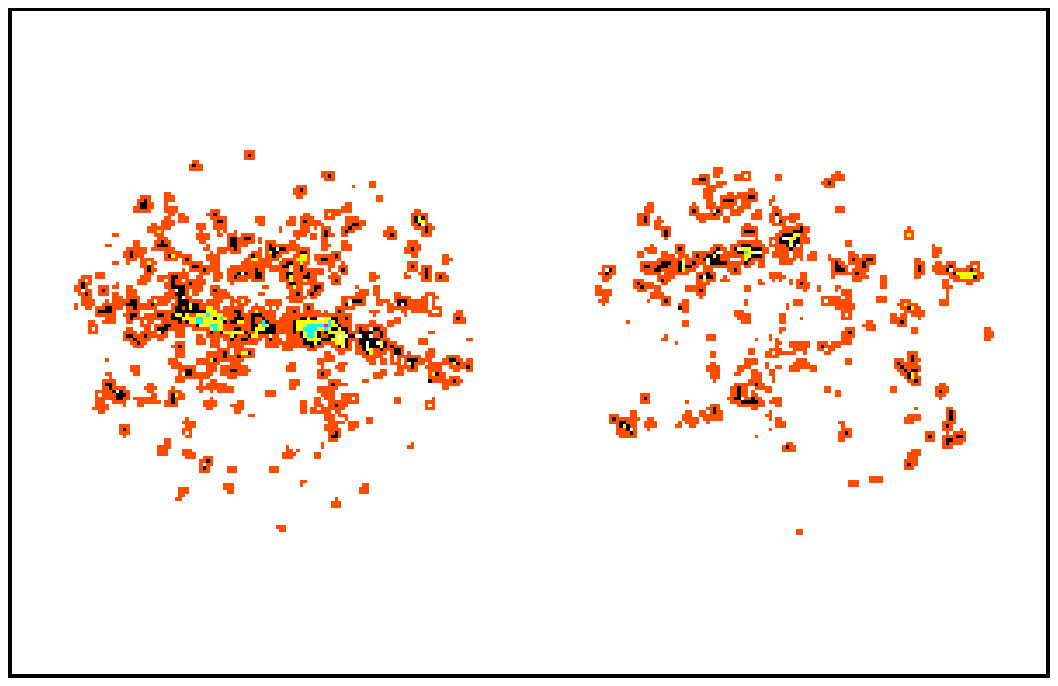}{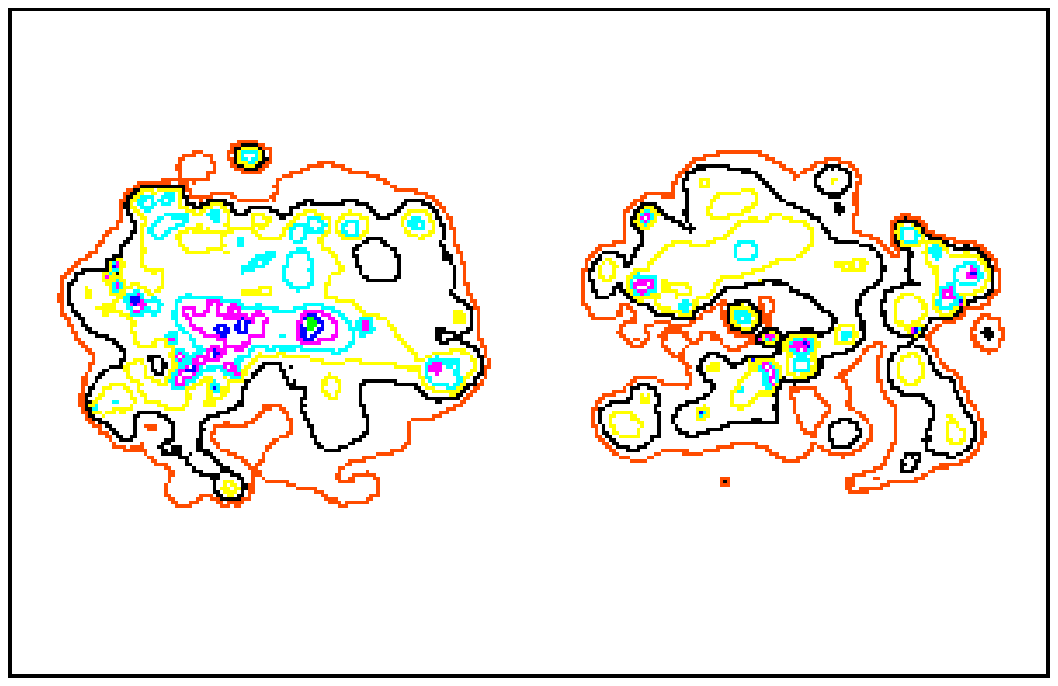}
\figcaption{
(a) left panel: Same with Figure 1b, but for $E \geq 10^{20.0}$ eV.
(b) right panel: Same with Figure 2a, but for $B = 100$ nG.
\label{fig2}}
\end{figure}

Next, the parameter dependence of these images is investigated.
In the left panel of Figure~\ref{fig2}, the image in the same
case with Figure~\ref{fig1}b but for $E \geq 10^{20.0}$ eV is shown.
We find that the image in Figure~\ref{fig2}a is more sharp and similar
to the image of the distribution of galaxies than that in
Figure~\ref{fig1}b. It is easily understood
because more energetic charged CRs propagates more straightly. 
In the right panel of Figure~\ref{fig2}, the image in the same
case with Figure~\ref{fig2}a but for $B = 100$ nG is shown.
Even for the energetic UHECRs, the deflection angle becomes relatively
large and the image is much distorted as long as the amplitude of the
extra-galactic magnetic field is taken to be large ($\sim 100$ nG).

\begin{figure}
\epsscale{1.0}
\plottwo{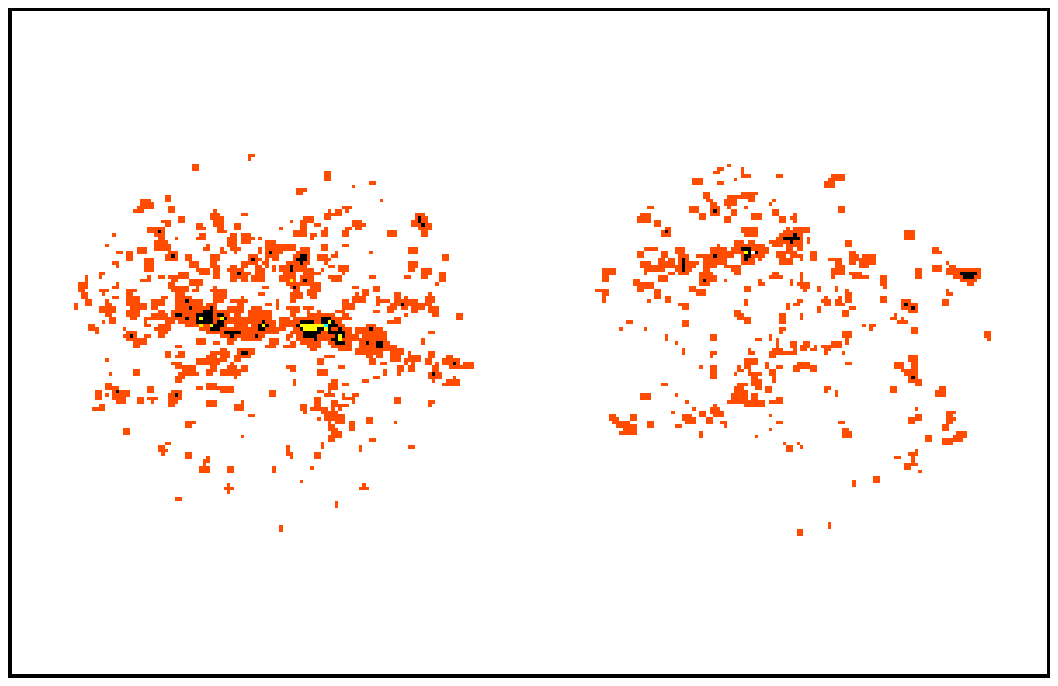}{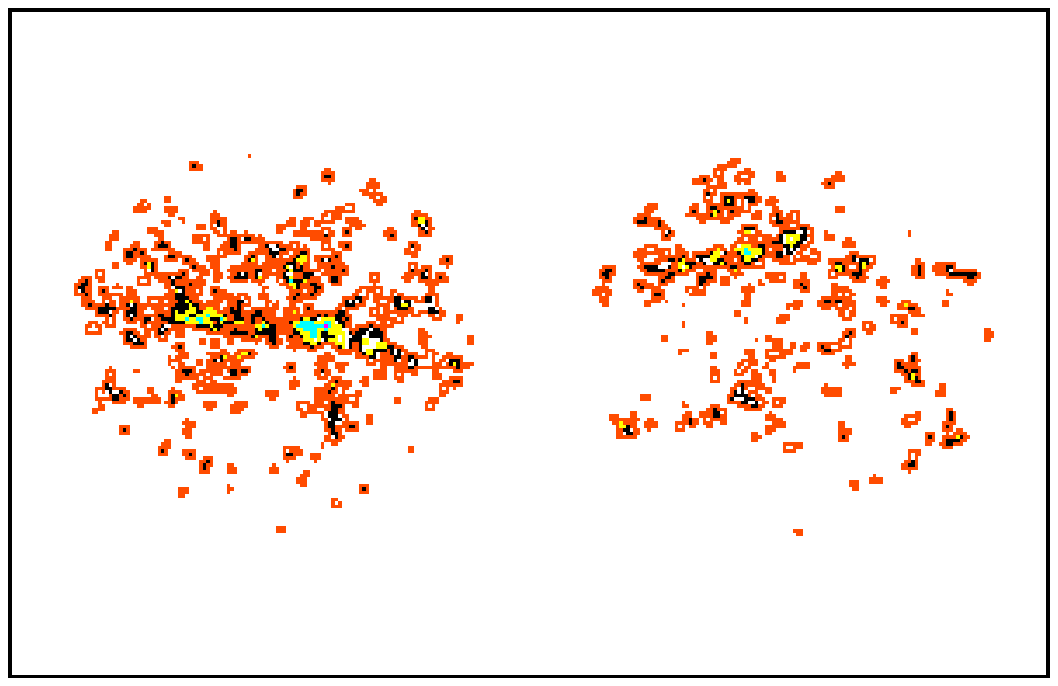}
\figcaption{
(a) left panel: Same with Figure 2a, but for $l_c = 1$ Mpc
(b) right panel: Same with Figure 2a, but for $l_c = 40$ Mpc
\label{fig3}}
\end{figure}

The dependence of these images on the correlation length,
$l_c$, is shown in Figure~\ref{fig3}. In the left panel of
Figure~\ref{fig3}, the image in the same case with Figure~\ref{fig2}a
but for $l_c = 1$ Mpc is shown. On the other hand, in the right panel of
Figure~\ref{fig3}, the image in the same case with Figure~\ref{fig2}a
but for $l_c = 40$ Mpc is shown. As is shown in Figure~\ref{fig3},
the effect of $l_c$ on the deflection is relatively weak. This is
because the deflection angle is proportional to $B$ and $l_c^{1/2}$,
respectively.

\placefigure{fig1}
\placefigure{fig2}
\placefigure{fig3}

\begin{figure}
\epsscale{1.0}
\plotone{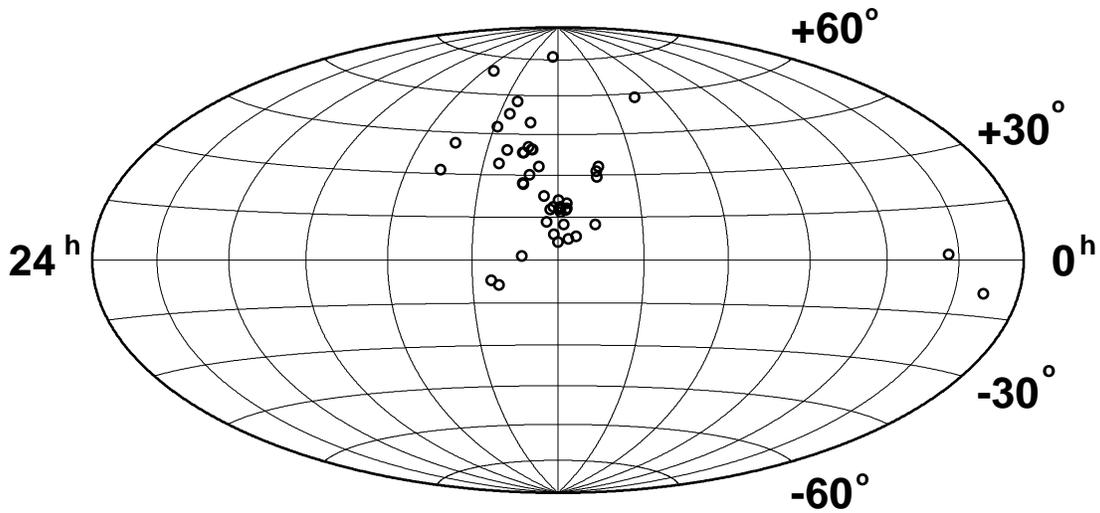}
\figcaption{
Arrival directions of UHECRs above 4$\times 10^{19}$ eV in equatorial
coordinates for the case of $B = 10$nG and $l_c = 1$Mpc.
The number of events is chosen to be 47 in order to compare
our results with the AGASA data (Hayashida 1999). The arrival directions
of UHECRs are restricted in the range $- 15^{\circ} \le b \le 80^{\circ}$
in order to compare our results with the AGASA data.
Clusters which mean the small-scale anisotropy of UHECRs can be found
clearly. 
\label{fig4}}
\end{figure}

\begin{figure}
\epsscale{1.0}
\plotone{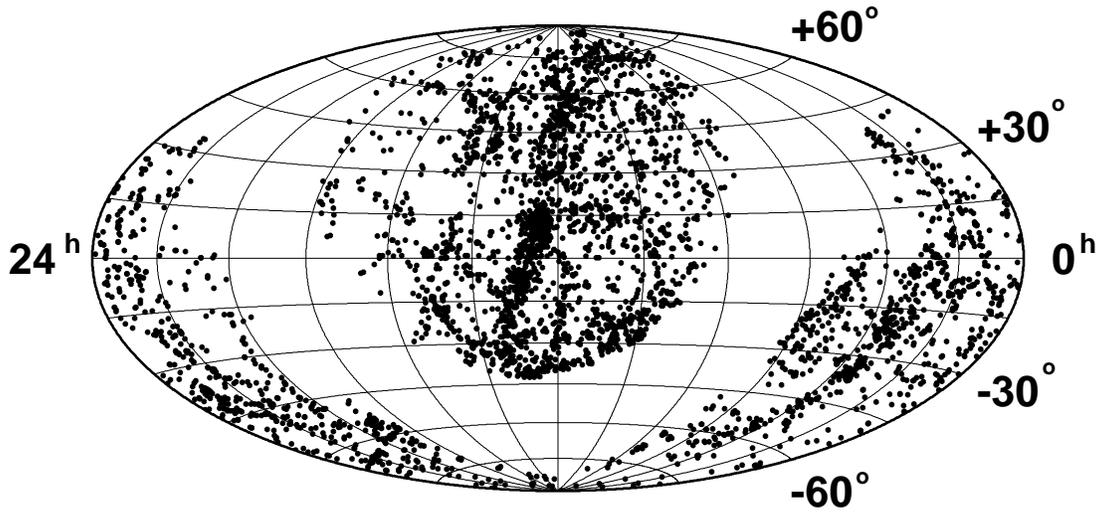}
\figcaption{
Angular distribution of galaxies in ORS data within 40Mpc in equatorial
coordinate. It is noted that this database does not contain any data
within $| b | \le 20^{\circ}$.
\label{fig5}}
\end{figure}

\begin{figure}
\epsscale{1.0}
\plotone{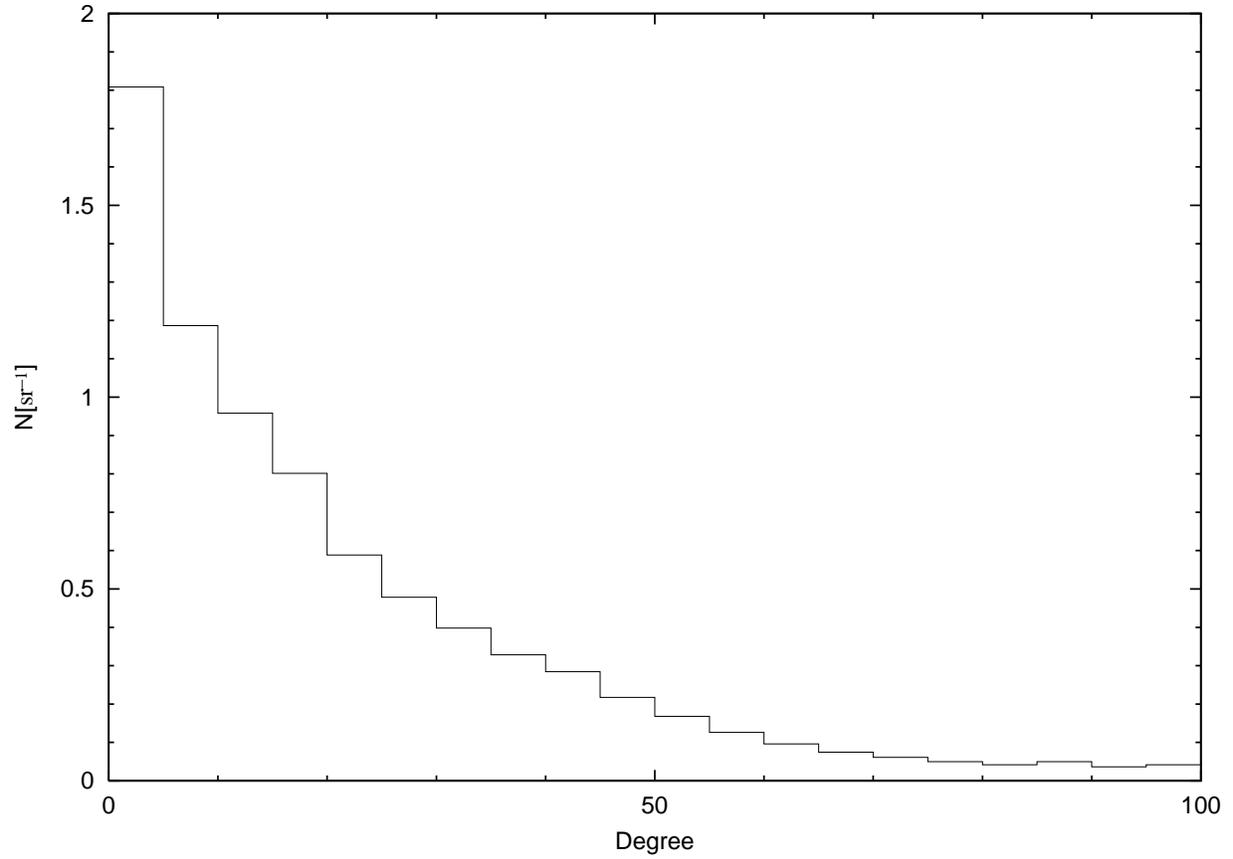}
\figcaption{
Distribution of the neighbor event ($\rm sr^{-1}$) as a function of
the separation angle between the two events
whose energies are above $4 \times 10^{19}$ eV.
The number of events in a data sample is chosen to be 47.
In order to estimate the mean value of every bin, the number of data
samples is set to be 1000 under the same condition ($B$ = 10nG and $l_c$ =
1Mpc). 
\label{fig6}}
\end{figure}

\begin{figure}
\epsscale{1.0}
\plotone{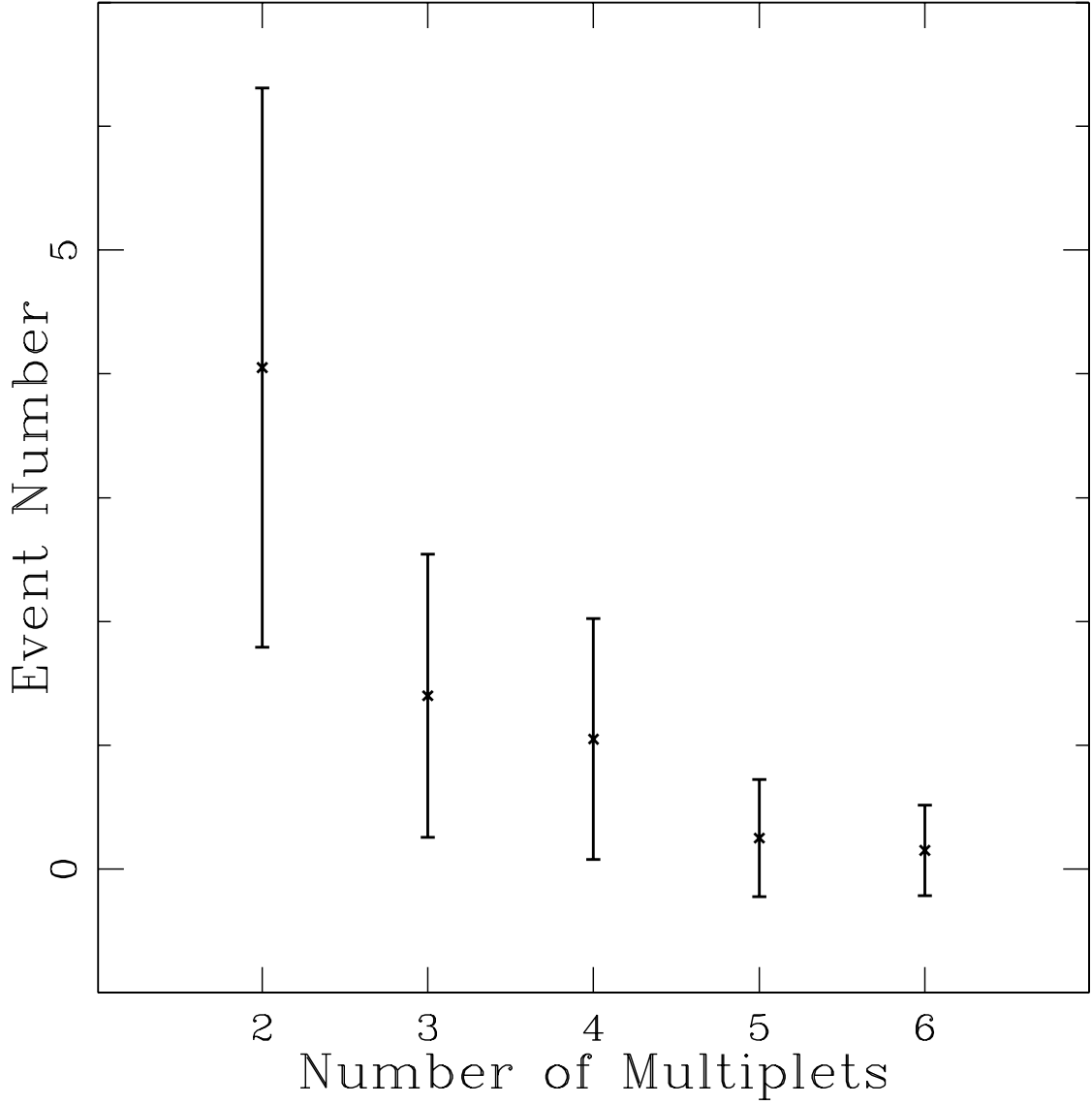}
\figcaption{
Number of clusters such
as doublets and triplets within a separation angle of $2.5^{\circ}$
in a data sample which contains 47 events, whose energies are above
$4 \times 10^{19}$ eV. The number of data samples is set to be 20
in order to obtain the mean values.
\label{fig7}}
\end{figure}

In Figure~\ref{fig4}, we show one realization of arrival directions of
UHECRs above 4$\times 10^{19}$ eV in equatorial coordinates for the case
of $B = 10$nG and $l_c = 1$Mpc. For comparison, we show the distribution
of the galaxies within 40 Mpc in ORS data catalogue in equatorial coordinates
in Figure~\ref{fig5}.
In Figure~\ref{fig4}, the number of events is chosen to be 47 in order
to compare
our results with the AGASA data (\cite{takeda99}). The arrival directions
of UHECRs are restricted in the range $- 15^{\circ} \le b \le 80^{\circ}$
in order to compare our results with the AGASA data.
Clusters which mean the small-scale anisotropy of UHECRs can be found
clearly. This result suggests that the probability of observing these
clusters is relatively high in our model. This is consistent with the
conclusion presented by Lemoine et al. (1999). In fact, we show in
Figure~\ref{fig6} the distribution of the neighbor event ($\rm
sr^{-1}$) as a function of the separation angle between the two events
whose energies are above $4 \times 10^{19}$ eV. It can be written as
\begin{eqnarray}
N ( \theta ) = \frac{1}{2 \pi | \cos \theta  - \cos (\theta + \delta \theta)
|} \sum_{ \theta
\le  \phi \le \theta + \delta \theta }  1 \;\;\; [ \rm  sr ^{-1} ],
\label{eqn100}
\end{eqnarray}
where $\phi$ denotes the separation angle of the two events.
$\delta \theta$ is taken to be $5^{\circ}$ in this analysis.
The number of events in a data sample is chosen to be 47.
The number of data samples is set to be 1000 in order to obtain the mean
value in every bin. We can find clearly the significant peak at samll
separation angles, which indicates the small-scale anisotropy of UHECRs.
We also show in Figure~\ref{fig7} the event number of clusters such
as doublets and triplets within a separation angle of $2.5^{\circ}$
in a data sample which contains 47 events, whose energies are above
$4 \times 10^{19}$ eV. The number of data samples is set to be 20
in order to obtain the mean values. We can find that the average number
of clusters such as doublets and triplets is consistent with the
AGASA data (\cite{takeda99}).

\placefigure{fig4}
\placefigure{fig5}
\placefigure{fig6}
\placefigure{fig7}

\begin{figure}
\epsscale{1.0}
\plotone{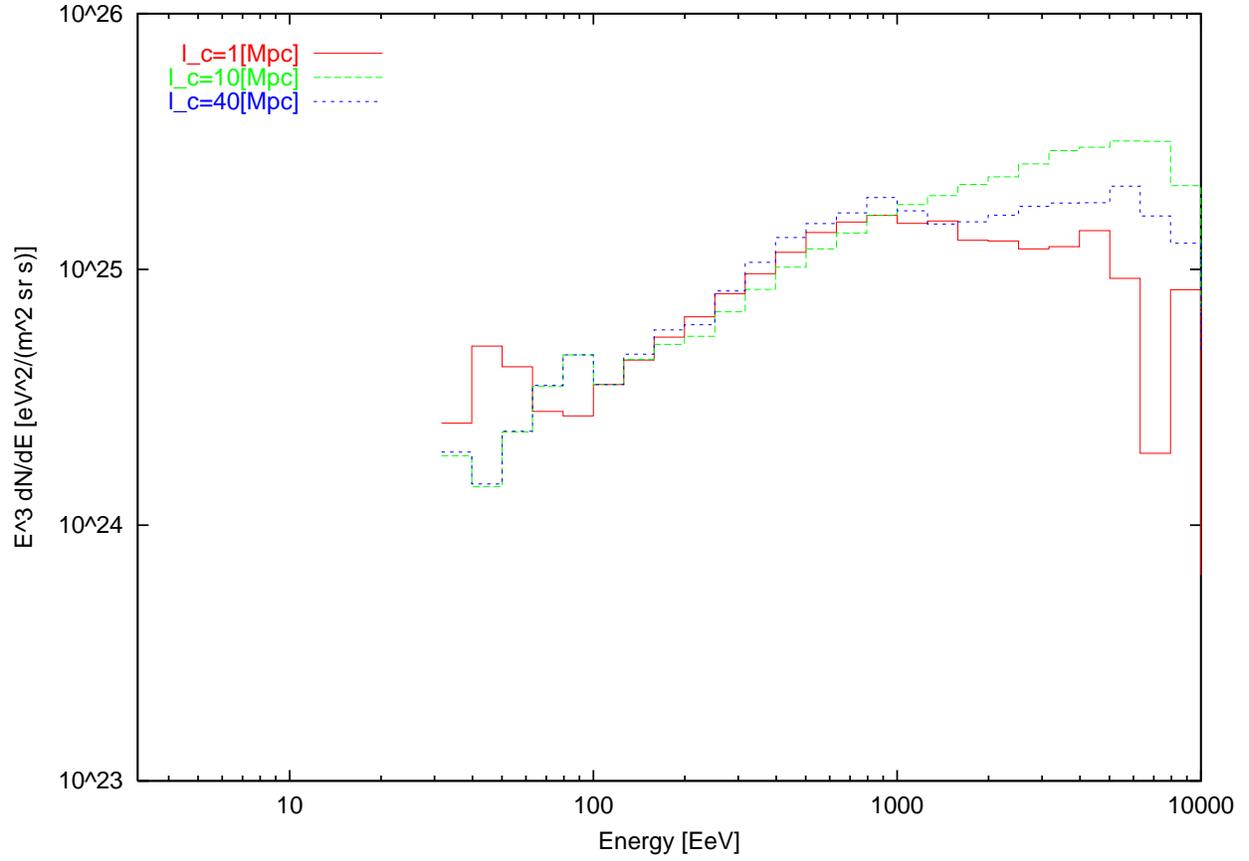}
\figcaption{
Calculated energy spectra of UHECRs. The parameters are set to be
$B=10$ nG and $H_0=75$ km/s/Mpc. The cases of $l_c$ = 1, 10 and 40 Mpc
are shown respectively. The energy spectra are normalized to the AGASA
data (Hayashida et al. 1999) at $E$ = 100 EeV.
\label{fig8}}
\end{figure}

Finally, the calculated energy spectra of UHECRs are shown in
Figure~\ref{fig8}. The parameters are set to be $B=10$ nG and
$H_0=75$ km/s/Mpc. The cases of $l_c$ = 1, 10 and 40 Mpc are shown
respectively. The energy spectra are normalized to the AGASA data
(\cite{takeda99})
at $E$ = 100 EeV. We think that the observed energy spectrum is well
reproduced by all the cases in Figure~\ref{fig8} without any fine-tuned
parameter.

\placefigure{fig8}

\subsection{Correlation Value}\label{statistics}
\indent

In this subsection, we discuss statistically the correlation between the
distribution of galaxies within GZK sphere and arrival directions of UHECRs.

\begin{figure}
\epsscale{1.0}
\plotone{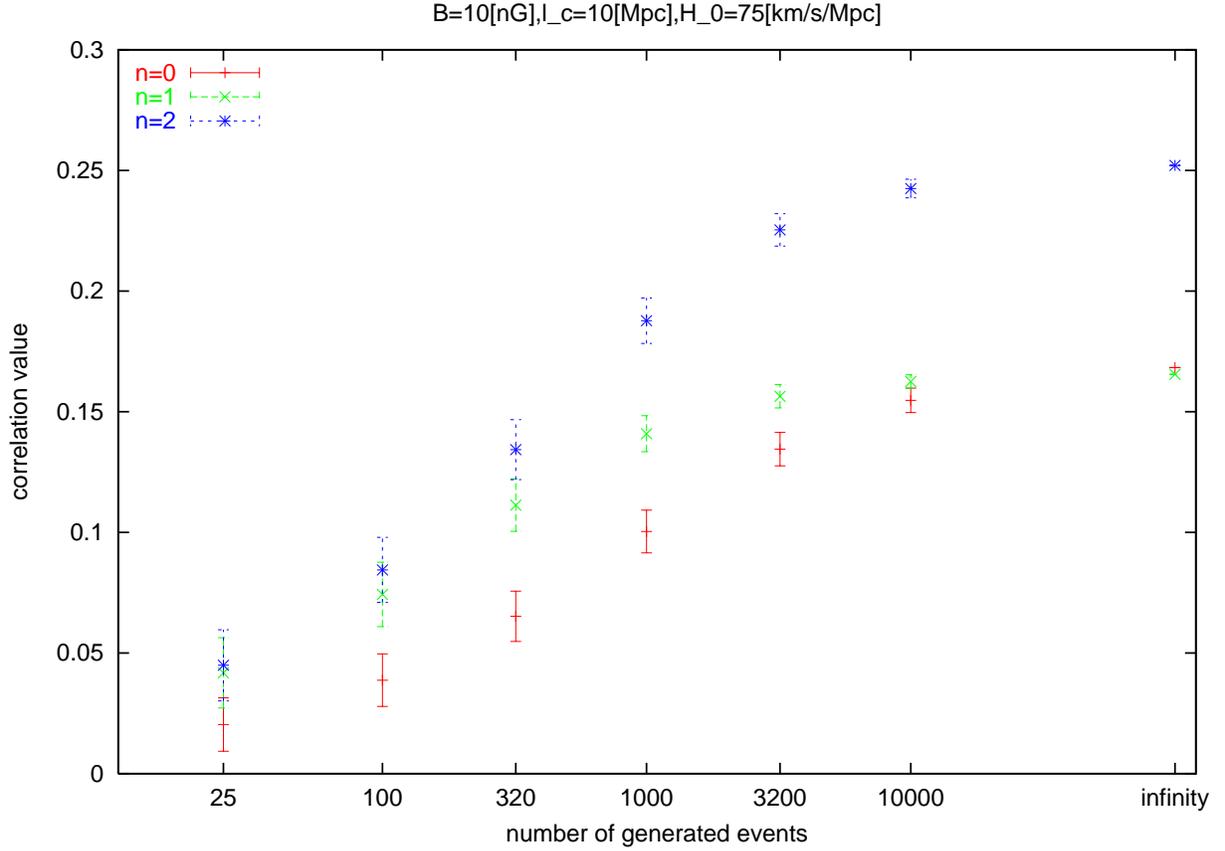}
\figcaption{
Correlation values for the case of $B=10$ nG, $l_c=10$ Mpc and $H_0=75.0$
km/s/Mpc indicating the influence of the selected energy range.
$n=0,$ $n=1,$ and $ n=2$ denote $E \geq 10^{19.5}$eV, $E \geq 10^{20.0}$eV,
and $E \geq 10^{21.0}$eV, respectively.
It is noted that the present observed number ($N$) of UHECRs of order
$10^1$ is too small to estimate the final correlation values. When $N$
is close to the order of $10^3$ -- $10^4$, the correlation values 
begin to converge.
\label{fig9}}
\end{figure}

\begin{figure}
\epsscale{1.0}
\plotone{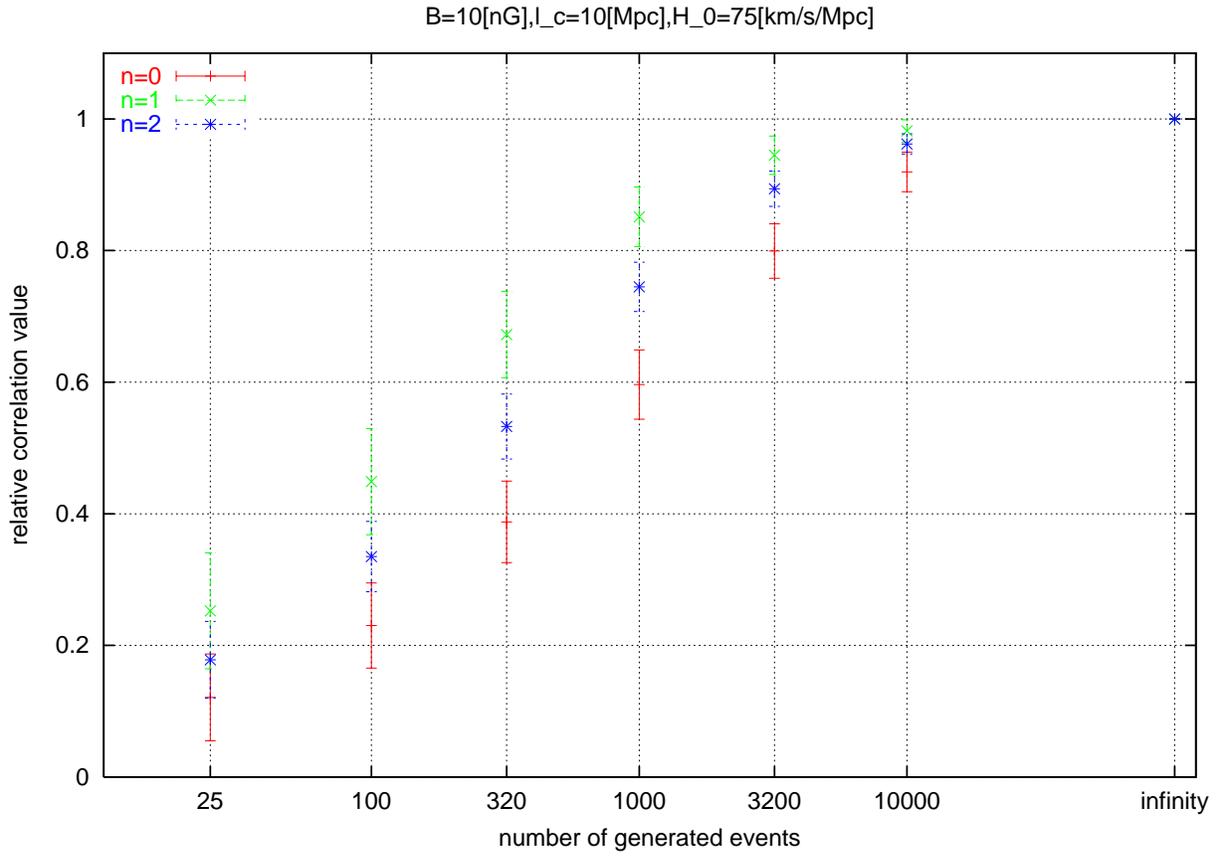}
\figcaption{
Normalized $\mit\Xi$ by its final value. The adopted parameters
are same with Figure 9.
\label{fig10}}
\end{figure}

At first, correlation values defined in
subsection~\ref{correlation} are shown in Figure~\ref{fig9} for the case of
$B=10$ nG, $l_c=10$ Mpc and $H_0=75.0$ km/s/Mpc indicating the influence
of the selected energy range. In this figure, $n=0$, $n=1,$ and $ n=2$
denote $E \geq 10^{19.5}$eV, $E \geq 10^{20.0}$eV, and $E \geq 10^{21.0}$eV,
respectively. It should be emphasized that the present
observed number ($N$) of UHECRs of order $10^1$ is too small to estimate
the final correlation values. When $N$ is close to the order of $10^3$ --
$10^4$, the correlation values begin to converge and final correlation
values can be estimated. It is well confirmed in Figure~\ref{fig10}, in which
each $\mit\Xi$ is normalized by its final value. We emphasize again that
a new generation of the ground-based large aperture experiments
is expected to detect about 1000 UHECRs until 2010 (\cite{zas01}). Moreover,
the mission EUSO is expected to detect about 1000 UHECRs per year
(\cite{bhattacharjee98}). So we will be able to determine the
source distribution in the very near future.

\placefigure{fig9}
\placefigure{fig10}

\begin{figure}
\epsscale{1.0}
\plotone{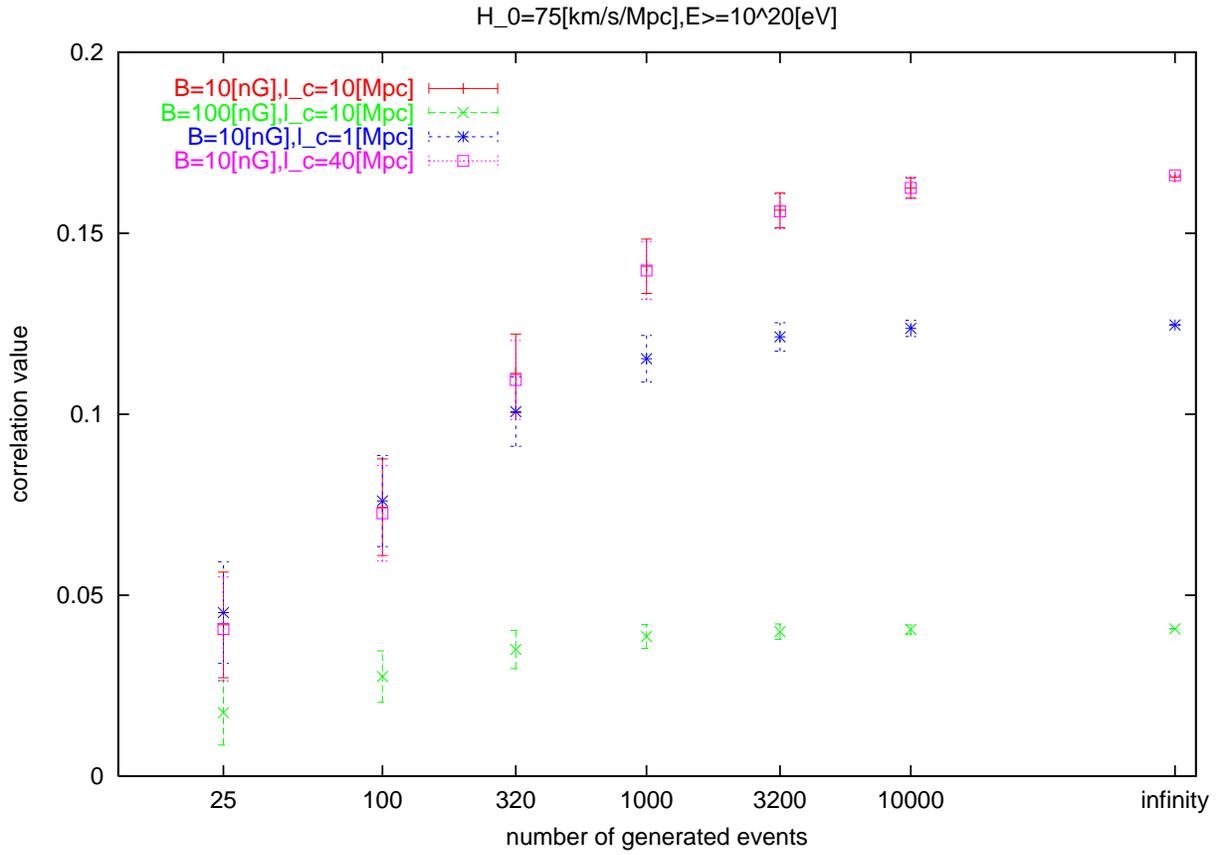}
\figcaption{
Correlation values in the case of $H_0=75$ km/s/Mpc and $E \geq 10^{20}$
eV indicating the influence of $B$ and $l_c$. It is confirmed that the
influence of $B$ is greater than $l_c$.
\label{fig11}}
\end{figure}

In Figure~\ref{fig11}, the correlation values are shown in the case of
$H_0=75$ km/s/Mpc and $E \geq 10^{20}$ eV indicating the influence of $B$
and $l_c$. It is confirmed again that the influence of $B$ is greater than
$l_c$, which is indicated in the previous subsection. Furthermore,
we can find again that the number of UHECRs of order $10^3$ is required
to estimate the final correlation value.

\placefigure{fig11}

\subsection{First and Second Harmonics}\label{firstsecond}
\indent

\begin{figure}
\epsscale{1.0}
\plottwo{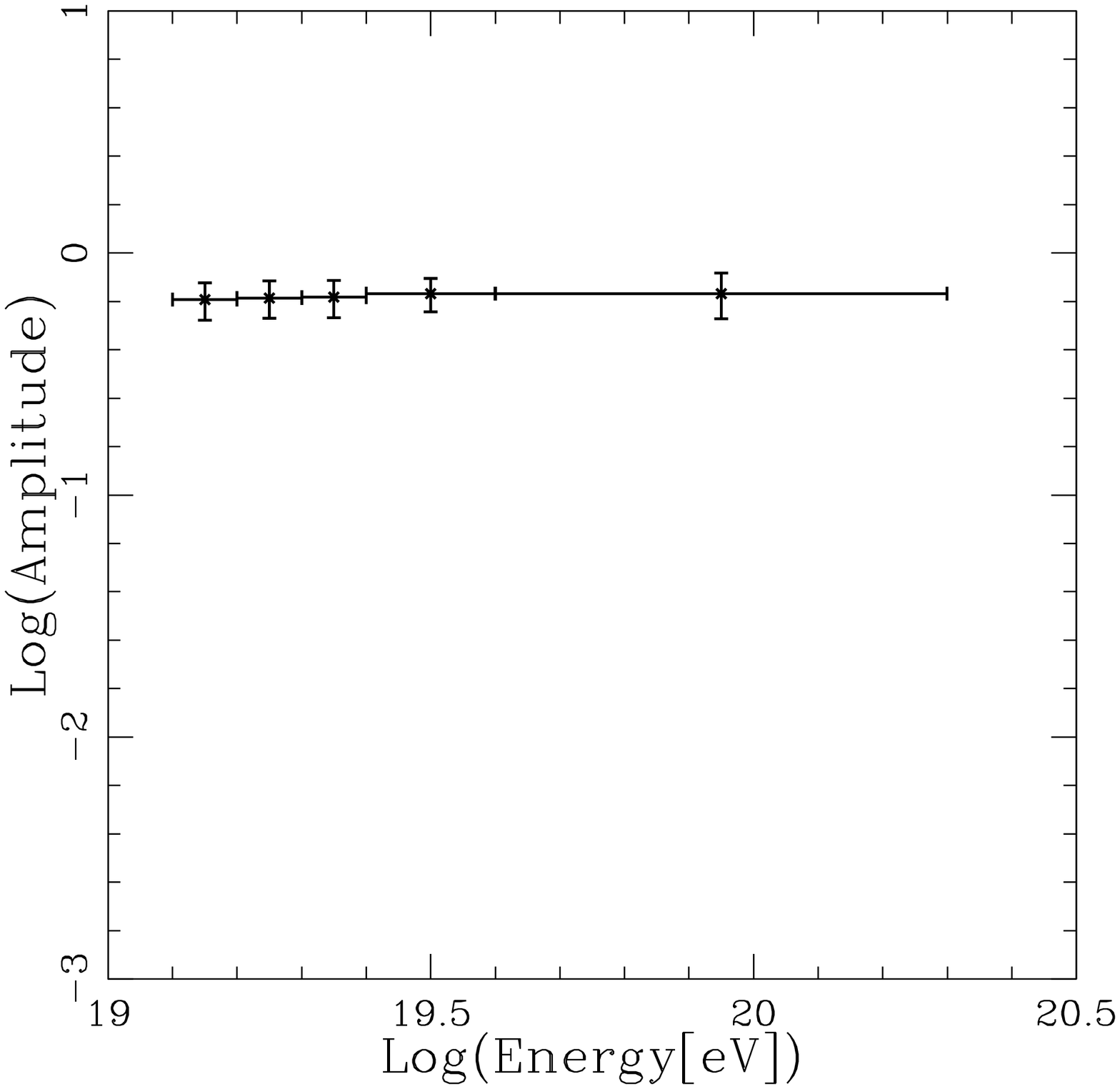}{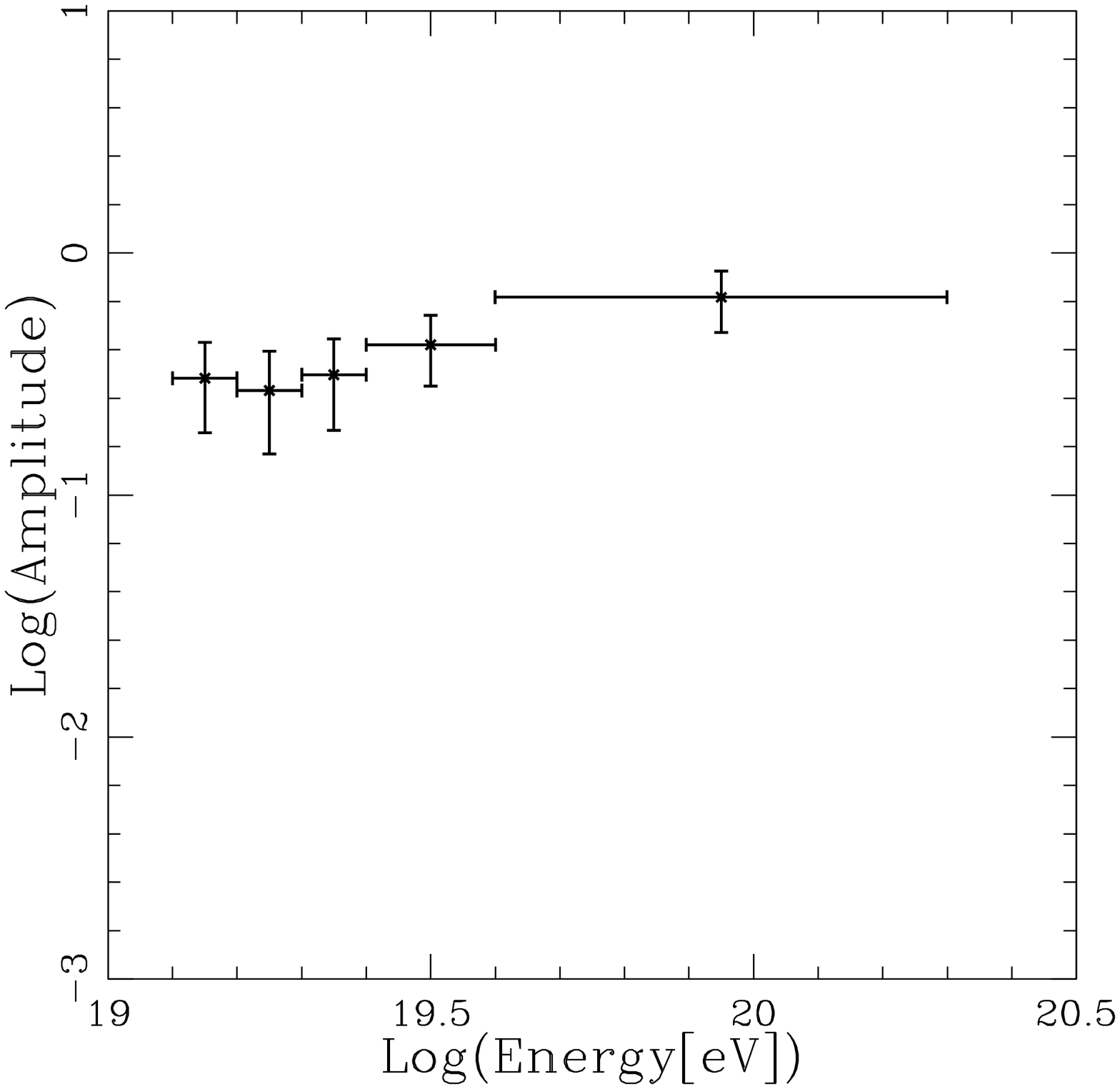}
\figcaption{
(a) left panel: amplitude of the first harmonics in galactic longitude in each
energy bin for the case of $B = 10$nG and $l_c = 1$Mpc.
$N$ is chosen to be 47 in the energy range (4$\times 10^{19}$
-- 2$\times 10^{20}$ eV) in order to compare our results with the AGASA
data (Hayashida 1999). The number of data samples is set to be 1000
in order to obtain the mean value and the standard deviation in every
bin. (b) right panel: same with Figure 12a, but for the second harmonics. 
\label{fig12}}
\end{figure}

\begin{figure}
\epsscale{1.0}
\plottwo{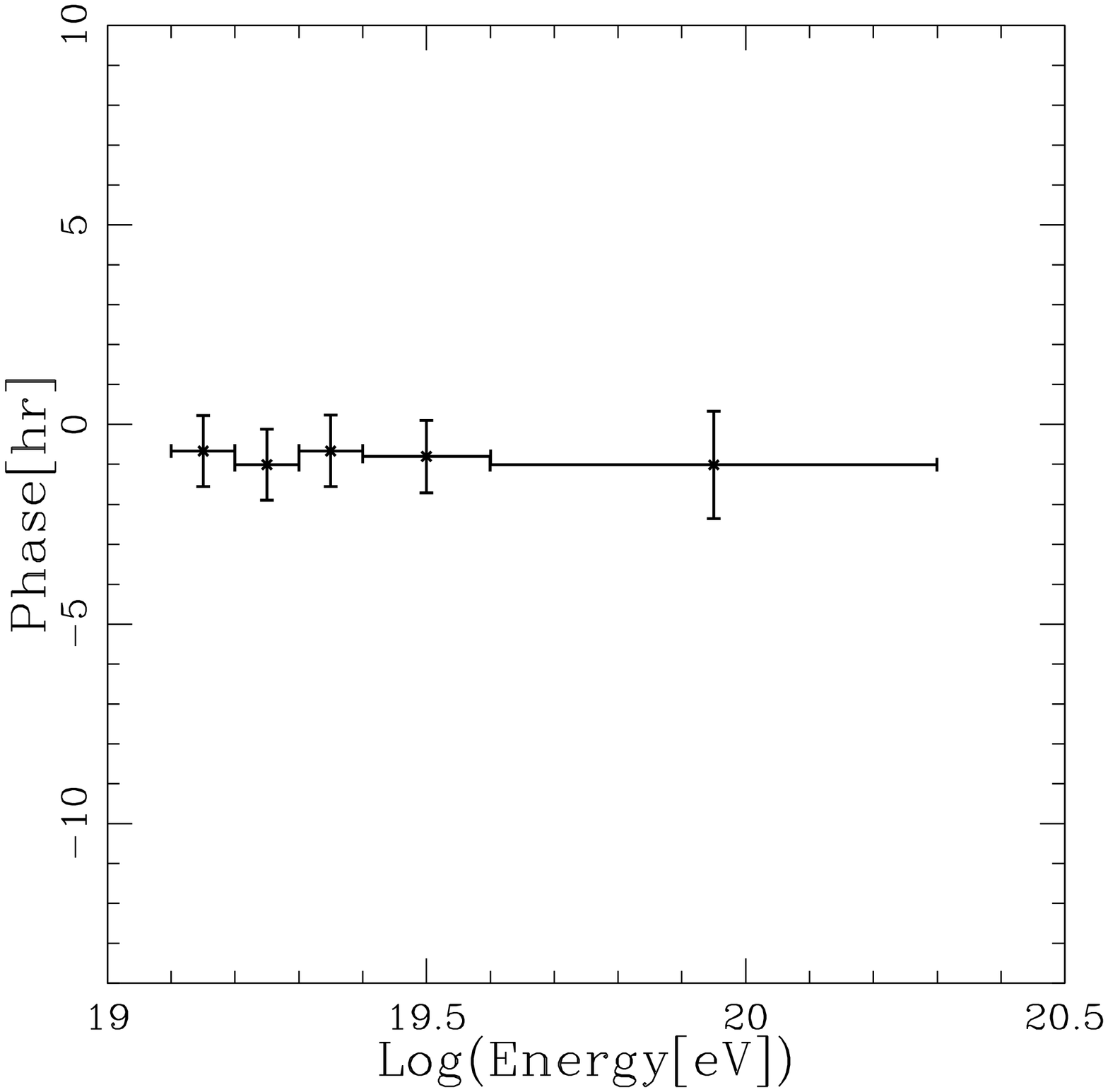}{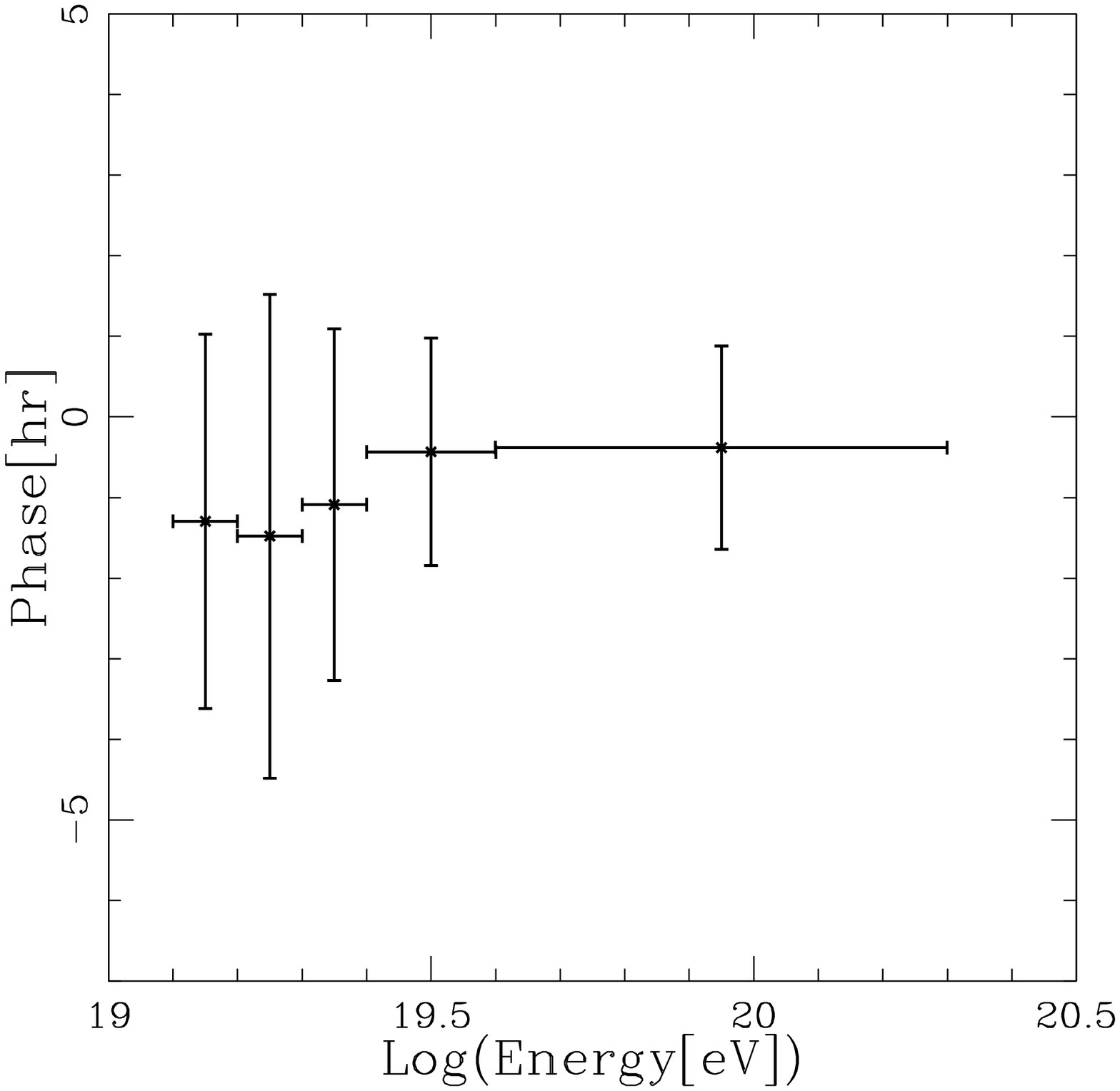}
\figcaption{
(a) left panel: same with Figure 12a, but for the phase of the first harmonics.
(b) right panel: same with Figure 13a, but for the second harmonics.  
\label{fig13}}
\end{figure}

\begin{figure}
\epsscale{1.0}
\plottwo{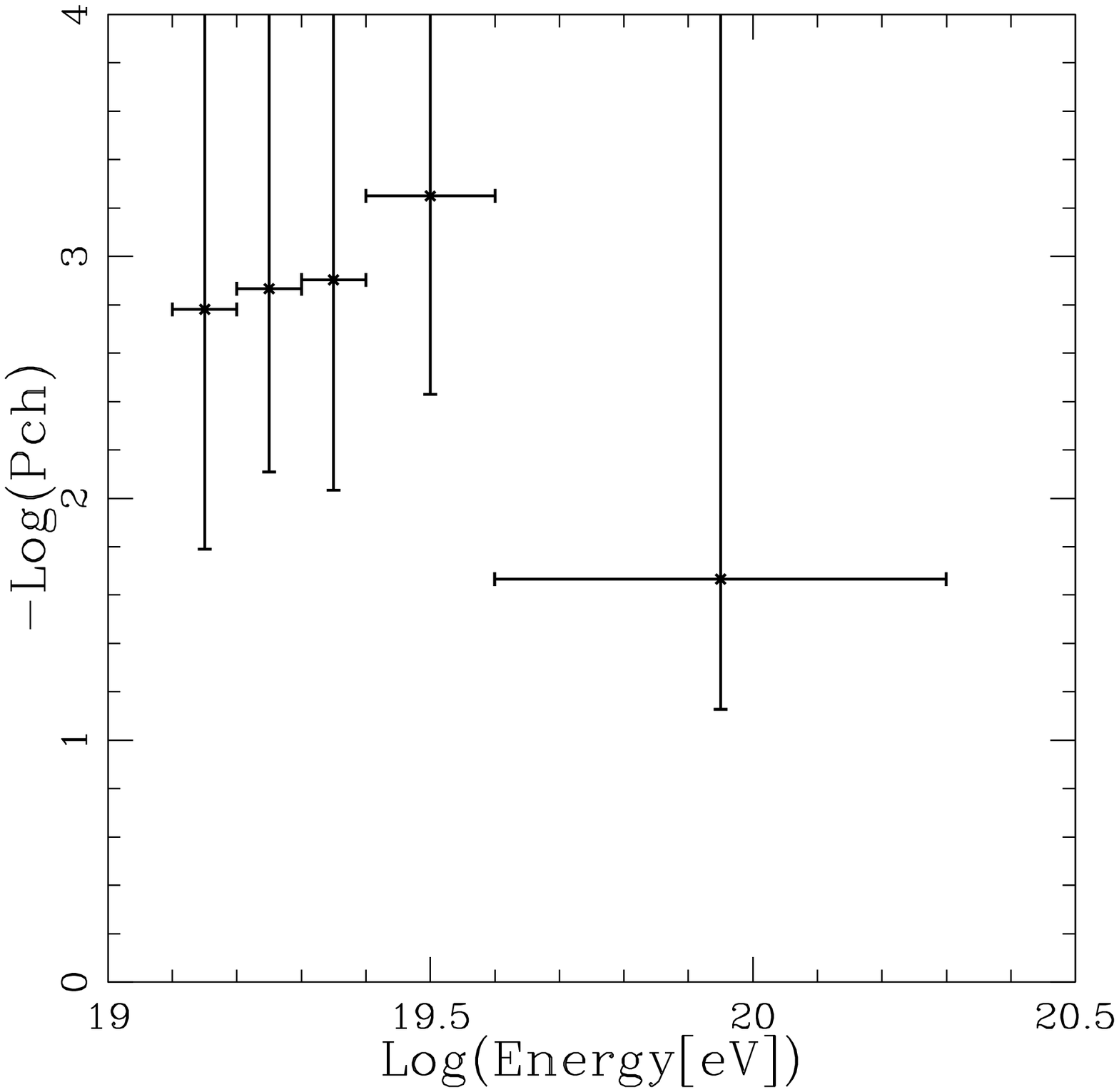}{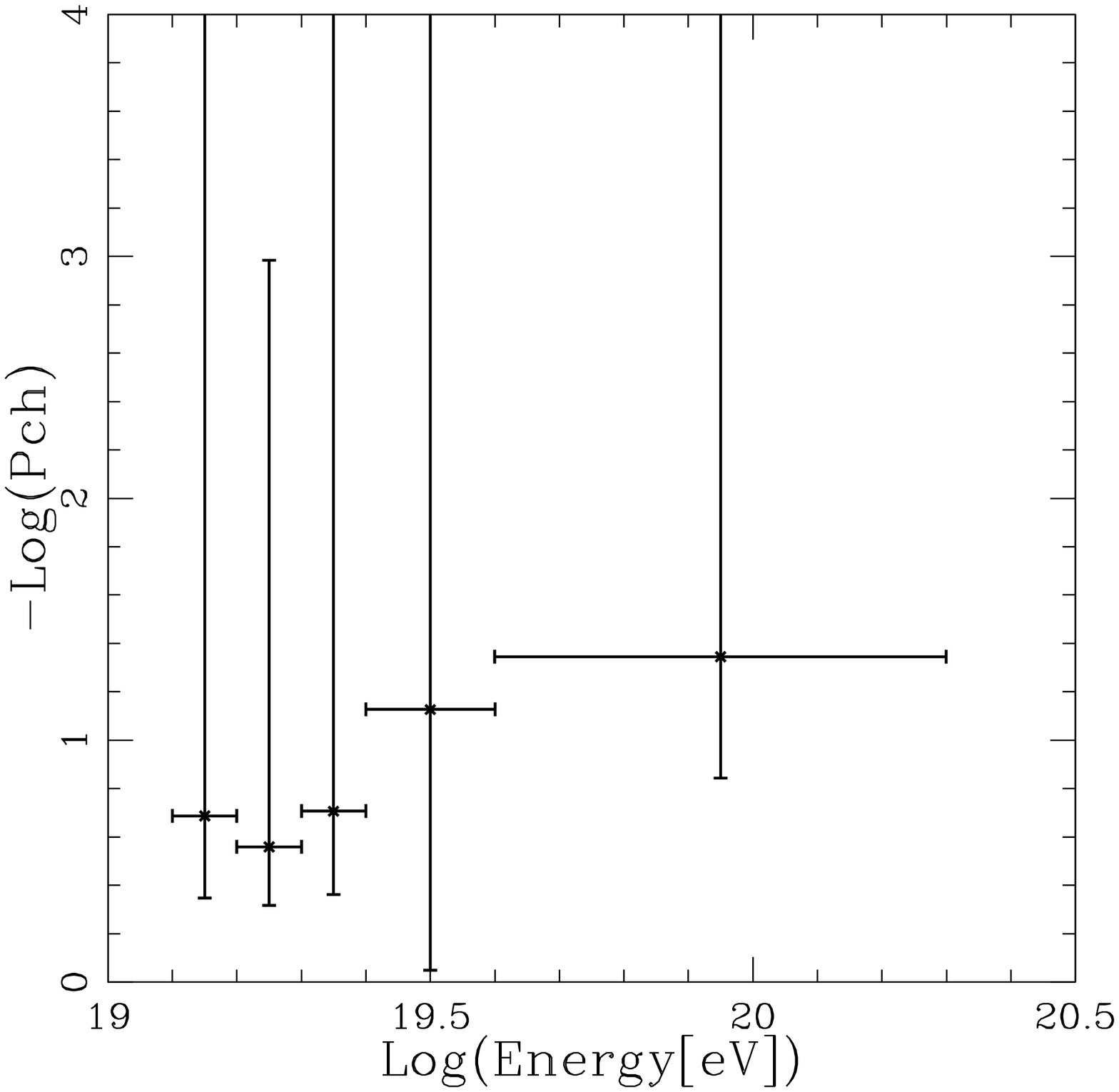}
\figcaption{
(a) left panel: same with Figure 12a, but for the chance probability of the
first harmonics. (b) right panel: same with Figure 14a, but for the second
harmonics.  
\label{fig14}}
\end{figure}

We show the results of the first (left panel) and second harmonics (right
panel) in the galactic longitude in Figure~\ref{fig12},~\ref{fig13},
and~\ref{fig14}.
The amplitudes, phases, and chance probability of harmonics
are shown in these figures. The parameters are set to be
$B = 10$nG and $l_c = 1$Mpc. The number of data, $n$, in one data sample
is chosen to be 47 in the energy range
(4$\times 10^{19}$ -- 2$\times 10^{20}$ eV) in order to compare our results
with the AGASA
data (\cite{takeda99}). The number of data samples is set to be 1000
in order to obtain the mean value and the standard deviation in every bin.

Compared with the AGASA data, the significant anisotropy 
on the arrival directions of UHECRs can be seen.
This may originate from the fact that the results of harmonic
analysis for the energy range $E \le 10^{19.5}$eV are not correct.
This is because the initial energy of UHECRs is restricted to be in the
range of $10^{19.5} \le E \le 10^{22}$eV to save the CPU time. This may also
originate from the incompleteness of the ORS database. That is, this database
does not contain any data within $| b | \le 20^{\circ}$ (see
Figure~\ref{fig5}). Also,
this problem may be solved if the source distribution is slightly
changed. For example, this problem may be solved if we assume that
UHECRs come from some of the galaxies such as AGNs and radio galaxies.
We will investigate in detail the dependence of our conclusion presented
in this study on the source distribution in the forthcoming paper.

\section{SUMMARY AND DISCUSSION} \label{summary}
\indent

We have performed the detailed numerical simulations on the propagation of
the UHE protons in the energy range $E=(10^{19.5} - 10^{22.0}$) eV in
the relatively strong extra-galactic magnetic field with strength
$B= (10, 100)$ nG within about 40 Mpc. In this case, the deflection
angles of UHECRs become so large that the no counterparts problem is simply
solved.

We have made the images of the angular distribution of UHECRs assuming that
the source distribution is proportional to the density distribution of matter,
or the galaxy distribution of the ORS data calatog within 40 Mpc.
It is noted that the images obtained by numerical simulations
do not mean the exact distributions of the UHECRs which we will detect indeed,
but the probability density distributions for the arrival directions of
UHECRs. We have found that the influence of $l_c$ on these images are very
weak, while that of $B$ is very strong.

We have found many clusters, which mean the small-scale anisotropy,
in our simulations. This is the advantage to assume that the source
distribution is not isotropic. It has been also shown that the observed
energy spectrum is well reproduced in our models without any fine-tuned
parameter.

We have used the correlation value $\mit\Xi$ in order to investigate
statistically the similarity between the distribution of arrival directions
of UHECRs and that of galaxies within the GZK sphere. We have found that
the values of the parameters are indistinguishable at the level of the
present number of the events, 25. Also, we have found that we are not
still able to determine whether the source distribution of UHECRs is
correlated with the SGP or not.

When the number of the detected events becomes ${\large O}(10^3)$, 
the correlation values for each parameter set begins to separate and converge. 
Thus, when the experimental data accumulate in the near future,
the estimation for the values of parameters will be possible by the
analysis of the correlation values. The expected number counts
by the experiment of the next generation such as TA, HiRes,
Auger, and EUSO are thought to be the order of $10^3$
(\cite{bhattacharjee98}). Thus, we will be able to determine the
source distribution and values of the parameters in this study
in the very near future.

Compared with the AGASA data (\cite{takeda99}), the significant anisotropy 
on the arrival directions of UHECRs can be seen in the analysis of
first and second harmonics. This may originate from the fact that the
results of harmonic
analysis for the energy range $E \le 10^{19.5}$eV are not correct.
This is because the initial energy of UHECRs is restricted to be in the
range of $10^{19.5} \le E \le 10^{22}$eV to save the CPU time.
This may also originate
from the incompleteness of the ORS database. That is, this database
does not contain any data within $| b | \le 20^{\circ}$. Also,
this problem may be solved if the source distribution is slightly
changed. For example, this problem may be solved if we assume that
UHECRs come from some of the galaxies such as AGNs and radio galaxies.
We will investigate in detail the dependence of our conclusion presented
in this study on the source distribution in the forthcoming paper.

\acknowledgements
This research has been supported in part by a Grant-in-Aid for the
Center-of-Excellence (COE) Research (07CE2002) and for the Scientific
Research Fund (199908802) of the Ministry of Education, Science, Sports and
Culture in Japan and by Japan Society for the Promotion of Science
Postdoctoral Fellowships for Research Abroad.


\end{document}